\documentclass[12pt]{article} \usepackage{a4wide,epsf}
  \renewcommand{\thefootnote}{\fnsymbol{footnote}}
  \newcommand{\D}{\stackrel{\leftrightarrow}{D}}
  \newcommand{\Dr}{\stackrel{\rightarrow}{D}}
  \newcommand{\Dl}{\stackrel{\leftarrow}{D}}
  \renewcommand{\theequation}{\arabic{section}.\arabic{equation}}

\begin{document}
  \begin{titlepage}
  \begin{flushright}
  \begin{tabular}{l}
  CERN--TH/96--12\\
  NORDITA--96--10--P\\
  hep-ph/9602323\\
  February 1995
  \end{tabular}
  \end{flushright}
  \vskip0.5cm
  \begin{center}
    {\Large \bf The $\rho$ Meson Light-Cone Distribution Amplitudes of
      Leading Twist Revisited\\ } \vskip1cm {\large Patricia Ball}
    \vskip0.2cm
    Theory Division, CERN  \\
    CH--1211 Geneva 23, Switzerland \\
    \vskip0.5cm {\large V.M.\ Braun}\footnote{On leave of absence from
      St.\ Petersburg Nuclear Physics Institute, 188350 Gatchina, Russia.}\\
    \vskip0.2cm
    NORDITA, Blegdamsvej 17, DK--2100 Copenhagen, Denmark\\
    \vskip2cm {\Large Abstract:\\[10pt]} \parbox[t]{\textwidth}{ We
      give a complete re-analysis of the leading twist quark-antiquark
      light-cone distribution amplitudes of longitudinal and
      transverse $\rho$ mesons.  We derive Wandzura-Wilczek type
      relations between different distributions and update the
      coefficients in their conformal expansion using QCD sum rules
      including next-to-leading order radiative corrections.  We find
      that the distribution amplitudes of quarks inside longitudinally
      and transversely polarized $\rho$ mesons have a similar shape,
      which is in contradiction to previous analyses.} \vskip2cm {\em
      Submitted to Physical Review D }
  \end{center}
  \end{titlepage}
  
  \renewcommand{\thefootnote}{\arabic{footnote}}
  \setcounter{footnote}{0}

  \section{Introduction}
  The theoretical interest in leading twist light-cone distribution
  amplitudes of hadrons is due to their r\^{o}le in the QCD
  description of hard exclusive processes \cite{BLreport}. In terms of
  the Bethe-Salpeter wave functions these distributions are defined by
  keeping track of the momentum fraction $x$ and integrating out the
  dependence on the transversal momentum $k_\perp$:
  \begin{equation}
     \phi(x) \sim \int_{k^2_\perp <\mu^2}d^2k_\perp\phi(x,k_\perp).
  \end{equation}  
  They describe probability amplitudes to find the hadron in a state
  with minimum number of Fock constituents and at small transverse
  separation (which provides an ultraviolet (UV) cut-off).  The
  dependence on the UV cut-off (scale) $\mu$ is given by
  Brodsky-Lepage evolution equations and can be calculated in
  perturbative QCD, while the distribution amplitudes at a certain low
  scale provide the necessary non-perturbative input for a rigorous
  QCD treatment of exclusive reactions with large momentum transfer
  \cite{exclusive}.
  
  Their investigation has been the subject of numerous studies.
  Chernyak and Zhitnitsky (CZ) have developed an approach to study
  moments of light-cone distributions using QCD sum rules
  \cite{CZreport}. Their main conclusion was \cite{CZ1,CZ2} that the
  pion and nucleon distribution amplitudes deviate strongly from the
  asymptotic distributions at large scales, which is a result still
  under debate.  Another result \cite{CZ1,CZZ1} was that the
  distribution amplitudes of longitudinally and transversely polarized
  $\rho$ mesons deviate from their asymptotic distributions in
  opposite directions: the longitudinal distribution is more wide
  while the transverse one is more narrow.  In the further discussion,
  the pion and nucleon distributions received most attention.
  
  The present paper is devoted to the re-evaluation of $\rho$ meson
  distributions along the lines of the approach of CZ and is mainly
  fuelled by newly emerged applications of light-cone distributions
  for the description of diffractive leptoproduction of vector mesons
  at HERA \cite{diffractive} and light-cone QCD sum rules for
  exclusive semileptonic $B\to\rho e\nu$ and radiative
  $B\to\rho\gamma$ weak decays \cite{ABS}.  The necessity of such an
  update is due to the following:
  
  First, the old calculations in \cite{CZ1,CZZ1} have used a very low
  normalization scale $\mu^2\sim 0.5\,$GeV$^2$ and a small value of
  the QCD coupling. Radiative corrections to the sum rules have been
  neglected.  With the larger values of $\alpha_s$ accepted nowadays
  the inclusion of the $O(\alpha_s)$ corections to the sum rules is
  mandatory.  The corresponding calculation is a new theoretical
  result of this paper.
  
  Second, there is a controversy in the sign of the contribution of
  four-fermion operators to the sum rule for the transverse vector
  meson as given by CZ \cite{CZZ1,CZreport}, and later calculations
  \cite{GRVW}. This sign difference apparently remained unnoticed and
  has dramatic consequences for the shape of the distribution.
  
  Third, earlier studies have not spent due attention to distributions
  of transversely polarized quarks in longitudinally polarized mesons.
  As first noted in \cite{ABS}, to leading twist accuracy these
  distributions are given in terms of longitudinal quark spin
  distributions.  We present a detailed derivation of the
  corresponding relations, the status of which is identical to that of
  the Wandzura-Wilczek relations \cite{WW} between the polarized
  nuleon structure functions $g_1(x,Q^2)$ and $g_2(x,Q^2)$.
  
  Our presentation is organized as follows.  In Sec.~2 we collect
  relevant definitions and give basic formulas for the expansion of
  the distribution amplitudes in contributions of conformal operators,
  which diagonalize the mixing matrix (Brodsky-Lepage kernels) to
  leading logarithmic accuracy. Section 3 is devoted to the analysis
  of QCD sum rules for the distributions in the transversely polarised
  $\rho$ meson, while Sec.~4 contains the sum rules for the
  longitudinally polarized $\rho$ meson. The final Sec.~5 contains a
  summary and some concluding remarks. We also include two appendices
  with the discussion of more technical issues.


  \section{The $\rho$ Meson Distribution Amplitudes}
  \setcounter{equation}{0}
  \subsection{Definitions}
  We define the light-cone distributions as matrix elements of
  quark-antiquark non-local gauge invariant operators at light-like
  separations \cite{CZreport}. For definiteness we consider the
  $\rho^+$ meson distributions; the difference to $\rho^0$ and
  $\omega$ is just a trivial isospin factor in the overall
  normalization.  The complete set of distributions to leading twist
  accuracy involves four wave functions \cite{ABS}:
  \begin{eqnarray}
  \langle 0 |\bar u(0)\sigma_{\mu\nu}d(x)
  |\rho^+(p,\lambda)\rangle & = &
  i(e^{(\lambda)}_\mu p_\nu -e^{(\lambda)}_\nu p_\mu)
  f_\rho^\perp \int_0^1 du\, e^{-iupx} \phi_\perp(u,\mu),
  \label{def1}\\
  \langle 0 |\bar u(0)\gamma_\mu d(x)
  |\rho^+(p,\lambda)\rangle &=& p_\mu \frac{(e^{(\lambda)} x)}{(px)}
  f_\rho m_\rho\int_0^1 du\, e^{-iupx} \phi_\parallel(u,\mu)
  \nonumber\\
   &&\mbox{}+\left( e^{(\lambda)}_\mu -p_\mu \frac{(e^{(\lambda)} x)}{(px)}
   \right) f_\rho m_\rho
   \int_0^1 du\, e^{-iupx} g_\perp^{(v)}(u,\mu)\,,
  \label{def2}\\
  \langle 0 |\bar u(0)\gamma_\mu\gamma_5 d(x)
  |\rho^+(p,\lambda)\rangle &=&
  -\frac{1}{4} \epsilon_{\mu\nu\rho\sigma} e^{(\lambda) \nu}
  p^\rho x^\sigma  f_\rho m_\rho
  \int_0^1 du\, e^{-iupx} g_\perp^{(a)}(u,\mu)\,,
  \label{def3}
  \end{eqnarray}
  where the gauge factors
  $$
  \mbox{Pexp}\left[ig\int_0^1 d\alpha\, x^\mu A_\mu(\alpha
    x)\right]
  $$
  are understood in between the quark fields.
  
  In the above definitions $p_\mu$ and $e^{(\lambda)}_\nu$ are the
  momentum and the polarization vector of the $\rho$ meson,
  respectively. The integration variable $u$ corresponds to the
  momentum fraction carried by the quark.  The normalization constants
  $f_\rho$ and $f_\rho^\perp$ (to be detailed later) are chosen in
  such a way that
  $$\int_0^1 du\, f(u)=1$$
  for all the four distributions $f=
  \phi_\perp , \phi_\parallel , g_\perp^{(v)} , g_\perp^{(a)}$.  The
  functions $\phi_\perp(u,\mu)$ and $\phi_\parallel(u,\mu)$ give the
  leading twist distributions in the fraction of total momentum
  carried by the quark in transversely and longitudinally polarized
  mesons, respectively.  The functions $g_\perp^{(v)}(u,\mu)$ and
  $g_\perp^{(a)}(u,\mu)$ describe transverse polarizations of quarks
  in the longitudinally polarized mesons and are to a large extent
  analogous to the spin structure function $g_2(x,Q^2)$ in polarized
  lepton-nucleon scattering. Similarly to the latter, they receive
  contributions of both leading twist 2 and non-leading twist 3, and
  the twist 2 contributions are related to the longitudinal
  distribution $\phi_\parallel(u,\mu)$ by Wandzura-Wilczek \cite{WW}
  type relations:
  \begin{eqnarray}
   g_\perp^{(v),{{\rm twist}\, 2}}(u,\mu) &=&
  \frac{1}{2}\left[
   \int_0^u dv \frac{\phi_\parallel(v,\mu)}{\bar v}
                 + \int_u^1 dv \frac{\phi_\parallel(v,\mu)}{ v}\right]\,,
  \nonumber\\
      g_\perp^{(a),{{\rm twist}\, 2}}(u,\mu) &=&
  2\left[  
   \bar u \int_0^u dv \frac{\phi_\parallel(v,\mu)}{\bar v}
                 + u \int_u^1 dv \frac{\phi_\parallel(v,\mu)}{ v}\right] \,.
  \label{WW1}
  \end{eqnarray}
  Here and below $\bar v \equiv 1-v$ etc.  Eq.~(\ref{WW1}) is derived
  in App.~A and presents one of our main results.
  
  The remaining twist 3 contributions to $g_\perp^{(v)}$,
  $g_\perp^{(a)}$ can be written in terms of three-particle
  quark-antiquark-gluon wave functions of transversely polarized
  vector mesons, cf.~\cite{CZreport,CZZ2}, and will not be considered
  in this paper. {}From now on we will drop the superscript ``twist
  2'', which is always implied.
  
  For some applications it is more convenient to rewrite (\ref{def2})
  as
  \begin{eqnarray}
  \langle 0 |\bar u(0)\gamma_\mu d(x)
  |\rho^+(p,\lambda)\rangle &=& p_\mu (e^{(\lambda)} x)
  f_\rho m_\rho\int_0^1 du\, e^{-iupx} \Phi_\parallel(u,\mu)
  \nonumber\\
  &&\mbox{}+ e^{(\lambda)}_\mu f_\rho m_\rho
   \int_0^1 du\, e^{-iupx} g_\perp^{(v)}(u,\mu)\,,
  \label{def2a}
  \end{eqnarray}
  introducing a new distribution function
  \begin{eqnarray}
    \Phi_\parallel(u,\mu) = 
  \frac{1}{2}\left[\bar u
   \int_0^u dv \frac{\phi_\parallel(v,\mu)}{\bar v}
                 - u \int_u^1 dv \frac{\phi_\parallel(v,\mu)}{ v}\right]\,.
  \label{WW2}
  \end{eqnarray}
  Eq. (\ref{WW2}) follows directly from (\ref{WW1}) and (\ref{def2a})
  by integration by parts.

  \subsection{Conformal Expansion and Renormalization}
  The separation between the quark and the antiquark in
  Eqs.~(\ref{def1})--(\ref{def3}) is assumed to be light-like, i.e.\ 
  $x^2=0$.  Extracting the leading behaviour of the matrix elements on
  the light-cone one encounters UV divergences, whose regularization
  yields a non-trivial scale-dependence which can be described by
  renormalization group methods \cite{exclusive,BLreport}.  The
  conformal invariance of QCD at tree level implies that operators
  with different conformal spin do not mix with each other to leading
  logarithmic accuracy. For the leading twist distributions
  $\phi_\perp(u,\mu)$ and $\phi_\parallel(u,\mu)$ it follows that the
  coefficients $a_n$ of their expansion in Gegenbauer polynomials
  $C_n^{3/2}(x)$ \cite{BE} (that is in contributions of operators with
  definite conformal spin) are renormalized multiplicatively to that
  accuracy:
  \begin{eqnarray}
  \phi(u,\mu)&=&6 u (1-u) \left[1 
   + \sum\limits_{n=2,4,\ldots} a_{n}(\mu)  C^{3/2}_{n}(2u-1)\right],
   \nonumber\\
    a_n(\mu) &=& a_n(\mu_0)
  \left(\frac{\alpha_s(\mu)}{\alpha_s(\mu_0)}\right)^{(\gamma_{(n)}-
\gamma_{(0)})/(2\beta_0)}
  \label{wf1}
   \end{eqnarray}
   with $\beta_0=11 - (2/3) n_f$. The one-loop anomalous dimensions
   are \cite{GW}
  \begin{eqnarray}
  \gamma_{(n)}^\parallel &=& \frac{8}{3} 
  \left(1-\frac{2}{(n+1)(n+2)}+4 \sum_{j=2}^{n+1} 1/j\right),
  \nonumber\\
  \gamma_{(n)}^\perp &=& \frac{8}{3} \left(1+4 \sum_{j=2}^{n+1} 1/j\right)\,.
  \label{eq:1loopandim} 
  \end{eqnarray}
  
  The conformal expansion of the distributions $g_\perp^{(v)}$,
  $g_\perp^{(a)}$ is more complicated and was derived in \cite{ABS}
  using the approach of Refs.~\cite{O,BF2}. We do not repeat the
  result in this paper, since to leading twist accuracy these
  distributions are not independent functions, but can be expressed in
  terms of $\phi_\parallel(u,\mu)$. One finds:
  \begin{eqnarray}
   g_\perp^{(a)}(u,\mu) = 6u(1-u)\left[1+\frac{1}{6}\,a_2^\parallel(\mu)\,
  C^{3/2}_2(\xi)+\ldots\right],
  \nonumber\\
  g_\perp^{(v)}(u,\mu) = \frac{3}{4}\,(1+\xi^2)+
  \frac{3}{16}\,a_2^\parallel(\mu)\,(15\xi^4-6\xi^2-1)+\ldots\,,
  \nonumber\\
  \Phi_\parallel(u,\mu) = \frac{3}{2}\, u(1-u)\xi\left[
  1+\frac{1}{4}\,a_2^\parallel(\mu)\,(15\xi^2-11)+\ldots\right]\,.
  \label{eq:trans-result}
  \end{eqnarray}
  Here and below we use the notation $\xi=2u-1$ as shorthand. The
  leading contributions in (\ref{eq:trans-result}) agree with the
  ``asymptotic distributions'' that were derived in Ref.~\cite{CZZ2}
  by a different method, but erroneously identified as being of twist
  3.\footnote{It is worthwhile to note that these leading terms
    correspond to the sum of contributions of leading and
    next-to-leading conformal spin, see \cite{ABS}.}

  \subsection{Non-Perturbative Input}
  
  The decay constants $f_\rho$, $f_\rho^\perp$ and the coefficients
  $a_n$ in the Gegenbauer expansion (\ref{wf1}) are intrinsic hadronic
  quantities and must be determined either experimentally or by
  non-perturbative methods. In particular, the decay constant $f_\rho$
  is measured~\cite{pdgold,pdg}:
  \begin{equation}
  f_{\rho^\pm} = (195\pm 7)\,{\rm MeV},\qquad f_{\rho^0} = (216\pm
  5)\,{\rm MeV}.
  \label{fvect}
  \end{equation}
  For other quantities, most of the existing information comes from
  QCD sum rules.  In what follows we summarize and update these
  calculations, taking into account radiative corrections and
  resolving some discrepancies in earlier studies.


  \section{Transversely Polarized $\rho$ Mesons}
  \setcounter{equation}{0}
  \subsection{The Tensor Coupling}
  The normalization of the leading twist quark-antiquark distribution
  in the transversely polarized $\rho$ meson is determined by the
  tensor coupling $f_\rho^\perp$, defined by
  \begin{equation}
  \langle 0 |\bar u\sigma_{\mu\nu}d
  |\rho^+(p,\lambda)\rangle  = 
  i(e^{(\lambda)}_\mu p_\nu -e^{(\lambda)}_\nu p_\mu)
  f_\rho^\perp\,,
  \label{eq:tensorcoupling}
  \end{equation}
  which can be estimated by studying the correlation function of two
  tensor currents within the framework of QCD sum rules \cite{SVZ}.
  We refer the reader to the reviews \cite{ShifmanBook} and
  \cite{CZreport} for detailed explanations of the method; the latter
  reference deals specifically with the determination of distribution
  functions.  A somewhat troublesome point in studying $f_\rho^\perp$
  is that the tensor current also couples to the positive parity
  $J^{PC}=1^{+-}$ state $b_1(1235)$ \footnote{$B(1235)$ in old
    classification.} \cite{pdg}:
   \begin{equation}
  \langle 0 |\bar u\sigma_{\mu\nu}d
  |b_1^+(p,\lambda)\rangle  = 
i\,\epsilon_{\mu\nu}^{\phantom{\mu\nu}\alpha\beta}e^{(\lambda)}_\alpha
p_\beta f_{b_1}^\perp\,.
  \label{eq:positiveparity}
  \end{equation}
  The correlation function of two tensor currents thus contains two
  Lorentz structures:
  \begin{eqnarray}
   \Pi_{\mu\nu} &=&
  i\!\int\!\! d^4y\, e^{iqy} \,\langle 0|
  T[\bar u(y)\sigma_{\mu\xi}x^\xi d(y)\bar d(0)
  \sigma_{\nu\xi}x^\xi u(0)]|0\rangle
  \nonumber\\
  &=&
  \frac{1}{q^2}[(qx)(q_\mu x_\nu+q_\nu x_\mu)-(qx)^2g_{\mu\nu}]\Pi^-(q^2)
  \nonumber\\ 
  &&{}
  +\frac{1}{q^2}[(qx)(q_\mu x_\nu+q_\nu x_\mu)-(qx)^2g_{\mu\nu}
  -q^2x_\mu x_\nu]\Pi^+(q^2)\,.
  \label{eq:decompose}
  \end{eqnarray}
  To compactify the Lorentz structure we have contracted the
  correlation function in two indices by the light-like vector $x_\mu$
  \cite{CZreport}.  The $\Pi^\pm(q^2)$ were calculated in \cite{GRVW}
  and correspond to intermediate states with positive (negative)
  parity, respectively:
  \begin{equation}
   \Pi^{\mp}(q^2) = -\frac{1}{8\pi^2}q^2\ln\,\frac{-q^2}{\mu^2}\Bigg[
  1+\frac{\alpha_s}{3\pi}\Big(\ln\,\frac{-q^2}{\mu^2}+\frac{7}{3}\Big)\Bigg]
  -\frac{1}{24 q^2}\langle \frac{\alpha_s}{\pi}\,G^2\rangle
  +\frac{16 \pi}{9q^4}\langle\sqrt{\alpha_s}\bar q q\rangle^2
  \Bigg[\frac{4}{9}\pm 1\Bigg]\,,
  \label{eq:Pipm}
  \end{equation}
  where we used vacuum saturation for the contributions of
  four-fermion operators.

  The correlation function $\Pi^-(q^2)$ can be used to extract the
  value of $f_\rho^\perp$, see e.g.~\cite{GRVW}.  Note, however, that
  it has a higher dimension than the correlation function of vector
  currents \cite{SVZ}, since in the latter case current conservation
  allows one to include one power of $q^2$ in the Lorentz structure.
  The higher dimension significantly reduces the accuracy of the sum
  rule, as it increases its sensitivity to higher resonances and the
  continuum.  In addition, the sign of the four-quark contribution is
  reversed, which does not allow to get a stable sum rule for the
  $\rho$ meson mass in this case, see \cite{GRVW}.  To overcome this
  difficulty, Chernyak and Zhitnitsky suggested to sum contributions
  of opposite parities. Since one has to assume
  $$
  \Pi^+(0)+\Pi^-(0) = 0
  $$
  to avoid an unphysical singularity at $q^2=0$ in
  Eq.~(\ref{eq:decompose}), it is legitimate to write a dispersion
  relation for the structure
  \begin{equation}
  \frac{\Pi^-(q^2)+\Pi^+(q^2)}{q^2} =
   -\frac{1}{4\pi^2}\ln\frac{-q^2}{\mu^2}\Bigg[
  1+\frac{\alpha_s}{3\pi}\Big(\ln\frac{-q^2}{\mu^2}+\frac{7}{3}\Big)\Bigg]
  -\frac{1}{12 q^4}\langle \frac{\alpha_s}{\pi}\,G^2\rangle
  +\frac{128 \pi}{81q^6}\langle\sqrt{\alpha_s}\bar q q\rangle^2\,.
  \label{eq:Pt}
  \end{equation}  
  Chernyak and Zhitnitsky speculated \cite{CZreport} that the
  approximation of local duality for the continuum contributions may
  be satisfied with better accuracy in sum rules with summation over
  different parity contributions, and noted that an additional
  advantage of using (\ref{eq:Pt}) is that contributions of particular
  four-fermion operators that are suspected to violate vacuum
  saturation cancel identically in this case.  The price to pay is
  that the sum rule contains an additional contribution of the
  $b_1$(1235) meson; since its mass, however, is very close to the
  continuum threshold in the $\rho$ meson channel, one may expect that
  this contamination has a minor effect.
  
  One can thus write down several different sum rules for
  $f_\rho^\perp$, each of which has its own advantages and
  disadvantages, and their agreement indicates consistency of the
  approach.  Using (\ref{eq:Pt}) one obtains
  \begin{eqnarray}
  \lefteqn{e^{-m_\rho^2/M^2} (f_\rho^\perp)^2(\mu) +
  e^{-m_{b_1}^2/M^2} (f_{b_1}^\perp)^2(\mu)\ =}\nonumber\\
   &=& \frac{1}{4\pi^2}\int\limits_0^{s_0}\!\! ds\,e^{-s/M^2} \left(
  1 + \frac{\alpha_s}{\pi}\left[\frac{7}{9}+\frac{2}{3}\,
  \ln\,\frac{s}{\mu^2}\right]\right) 
  -\frac{1}{12M^2}\,\langle\frac{\alpha_s}{\pi}G^2\rangle
  -\frac{64\pi}{81M^4}\,\langle\sqrt{\alpha_s}\bar q
  q\rangle^2\,.\makebox[0.8cm]{}
  \label{eq:SRft1}
  \end{eqnarray}
  On the other hand, starting from the correlation functions
  $\Pi^\mp(q^2)$, one gets
  \begin{eqnarray}
  \lefteqn{
  m_\rho^2 e^{-m_\rho^2/M^2} (f_\rho^\perp)^2(\mu)  = }
  \nonumber\\
  &=&
  \frac{1}{8\pi^2}\int\limits_0^{s_0^\rho}\!\! sds\,e^{-s/M^2} \left(
  1 + \frac{\alpha_s}{\pi}\left[\frac{7}{9}+\frac{2}{3}\,
  \ln\,\frac{s}{\mu^2}
  \right]\right) +\frac{1}{24}\,\langle\frac{\alpha_s}{\pi}G^2\rangle
  +\frac{208\pi}{81M^2}\,\langle\sqrt{\alpha_s}\bar q
  q\rangle^2\,,\makebox[0.8cm]{}
  \label{eq:SRft2}\\
  \lefteqn{
  m_{b_1}^2e^{-m_{b_1}^2/M^2} (f_{b_1}^\perp)^2(\mu)  =}
  \nonumber\\
  &=&
  \frac{1}{8\pi^2}\int\limits_0^{s_0^{b_1}}\!\! sds\,e^{-s/M^2} \left(
  1 + \frac{\alpha_s}{\pi}\left[\frac{7}{9}+\frac{2}{3}\,
  \ln\,\frac{s}{\mu^2}
  \right]\right) +\frac{1}{24}\,\langle\frac{\alpha_s}{\pi}G^2\rangle
  -\frac{80\pi}{81M^2}\,\langle\sqrt{\alpha_s}\bar q
  q\rangle^2\,,\makebox[0.8cm]{}
  \label{eq:SRft3}
  \end{eqnarray}
  where $s_0^\rho\simeq 1.5$ GeV$^2$ \cite{SVZ} and $s_0^{b_1}\simeq
  2.3$ GeV$^2$ \cite{GRVW} are the continuum thresholds in the $\rho$
  and $b_1$ channels, respectively.  The continuum threshold $s_0$ for
  the ``mixed parity'' sum rule (\ref{eq:SRft1}) is discussed below.
  $M^2$ is the Borel-parameter.  Note that the sign of the
  contribution of four-fermion operators in (\ref{eq:SRft1}) is
  opposite to the result given in \cite{CZZ1,CZreport}.  We have
  recalculated this contribution and confirm the sign as obtained in
  \cite{GRVW}.
  
  In the numerical analysis we use $\alpha_s(\mu=1\,\mbox{\rm GeV})=
  0.56$, i.e.\ $\Lambda_{\overline{\rm MS}}^{(3)} = 0.4\,{\rm GeV}$,
  corresponding to the world average $\alpha_s(m_Z)=0.119$~\cite{pdg}.
  For the condensates we take the standard values \cite{SVZ}
  \begin{equation}
  \langle \frac{\alpha_s}{\pi}\,G^2\rangle = (0.012\pm 0.006)\,{\rm
    GeV}^4\,, \qquad \langle\sqrt{\alpha_s} \bar q q\rangle^2 =
  0.56\,(-0.25\,{\rm GeV})^6 \,.
  \end{equation}
  The sum rules and the couplings are evaluated at $\mu=1\,{\rm GeV}$.
  We have checked that changing the scale in the range
  $\mu^2=(1-2)\,{\rm GeV}^2$ does not have any noticeable effect,
  provided the extracted couplings are related by renormalization
  group scaling:\footnote{In this paper we stay consistently within
    $O(\alpha_s)$ accuracy and do not attempt a renormalization group
    improvement of sum rules (see e.g. \cite{BBBD}).}
  $$f^\perp(1\,{\rm GeV}) = \left[\frac{\alpha_s(1\,{\rm
        GeV})}{\alpha_s(\mu)}\right]^{4/27} f^\perp(\mu)\,.  $$

  \begin{figure}
    \vspace*{-10pt} \centerline{\epsffile{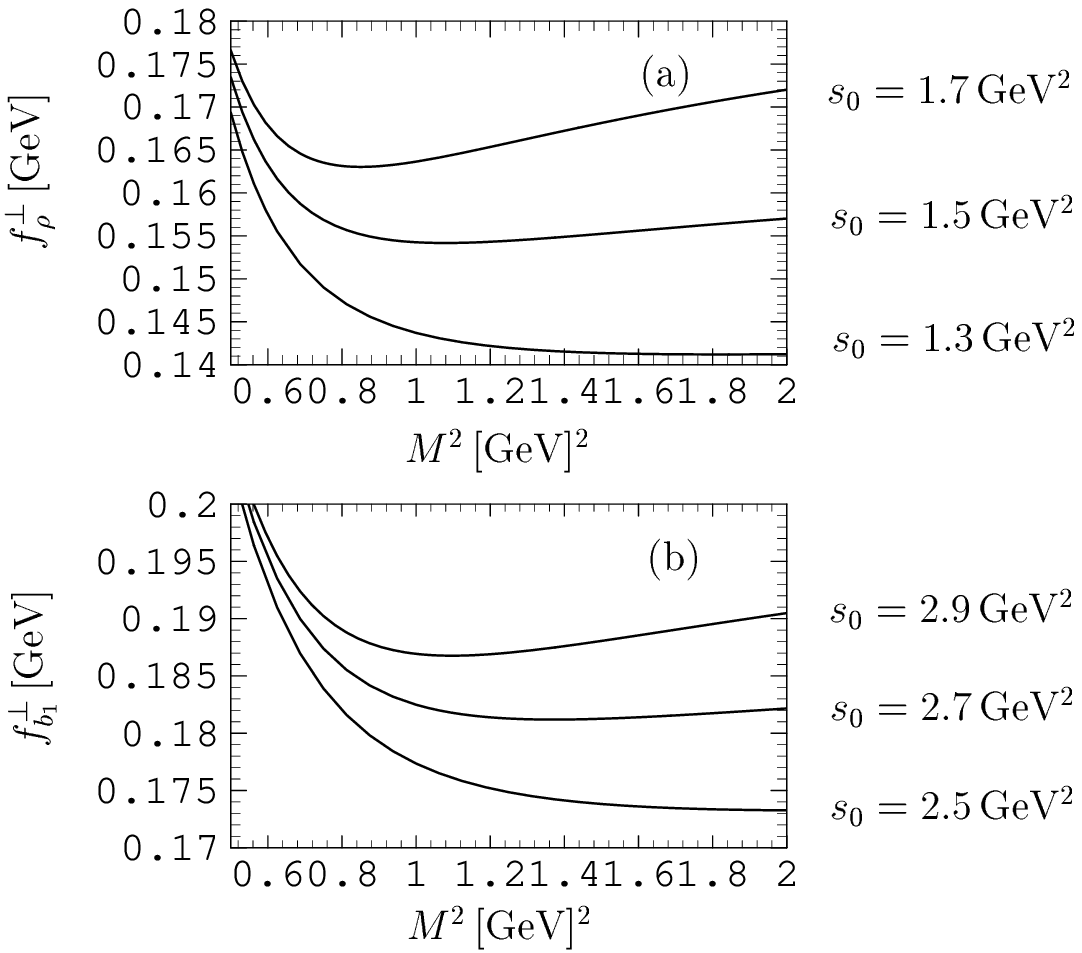}} \vspace*{-10pt}
  \caption[]{(a) $f_\rho^\perp(1\,{\rm GeV})$ from 
    Eq.~(\protect{\ref{eq:SRft2}}) as function of the Borel parameter
    $M^2$ for different values of the continuum threshold $s_0$. (b)
    The same for $f_{b_1}^\perp(1\,{\rm GeV})$ from
    Eq.~(\protect{\ref{eq:SRft3}}).  }\label{fig:pureparity}
  \vspace*{5pt} \centerline{\epsffile{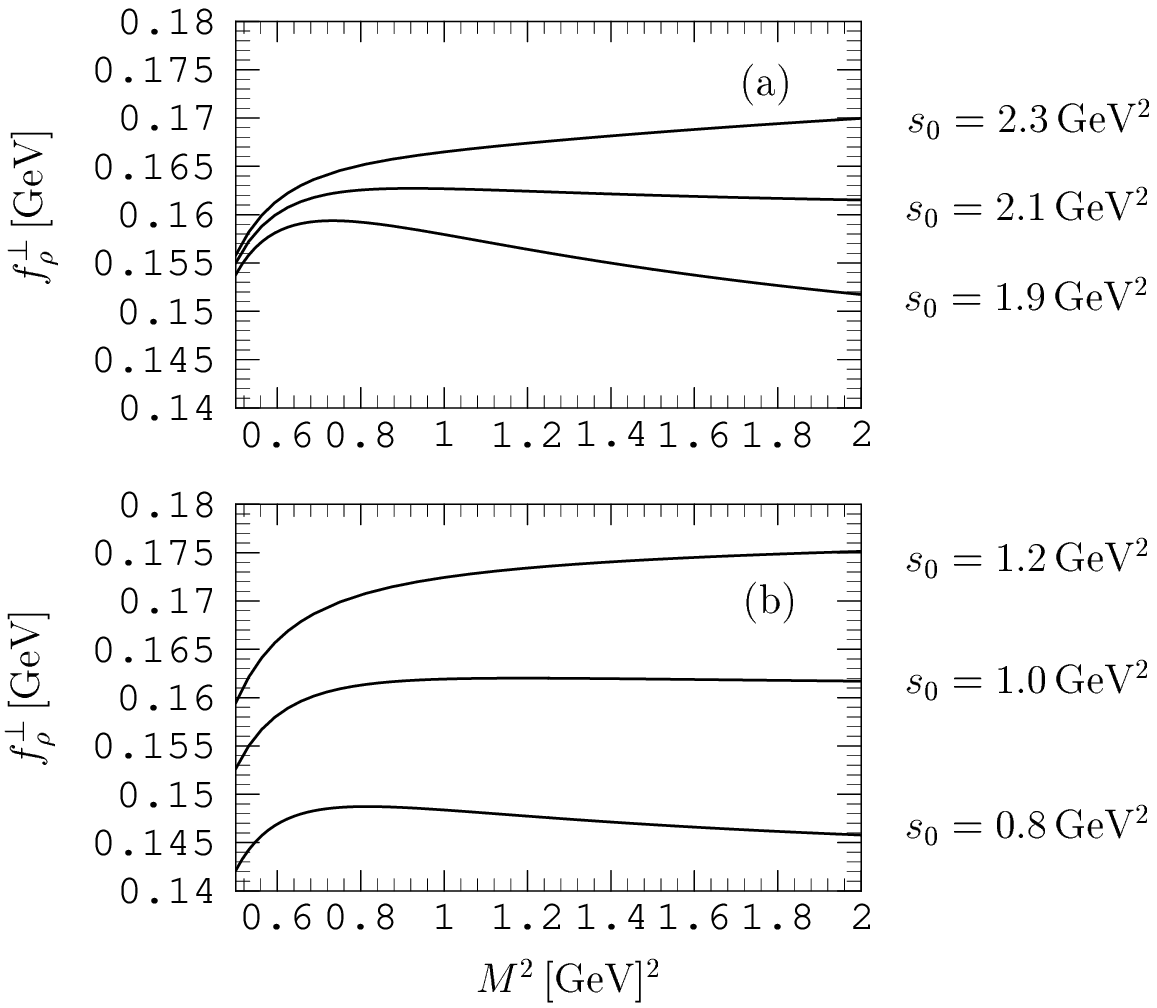}} \vspace*{-10pt}
  \caption[]{(a) $f_\rho^\perp(1\,{\rm GeV})$ from 
    Eq.~(\protect{\ref{eq:SRft1}}) as function of the Borel parameter
    $M^2$ for different values of the continuum threshold $s_0$.
    $f_{b_1}^\perp(1\,{\rm GeV})$ is put to 180$\,$MeV. (b) The same
    with the $b_1$ contribution put to the continuum, i.e.\ 
    $f_{b_1}^\perp=0.$ }\label{fig:mixedparity}
  \end{figure}
  We start with the ``pure parity'' sum rules in Eqs.~(\ref{eq:SRft2})
  and (\ref{eq:SRft3}). The values of the couplings extracted from
  these sum rules are shown in Fig.~\ref{fig:pureparity}a and
  Fig.~\ref{fig:pureparity}b as functions of the Borel parameter for
  several choices of the continuum thresholds.  Requiring best
  stability in the ``working window'' of the Borel parameter
  $1<M^2<1.5$~GeV$^2$, we find
  \begin{eqnarray}
  f_\rho^\perp = (160\pm 10) \,\mbox{\rm MeV},&& \qquad
  s_0^\rho =1.5\,\mbox{\rm GeV}^2\,,
  \label{frhot1}\\
f_{b_1}^\perp = 180\,{\rm MeV},&& \qquad s_0^{b_1} = 2.7\,{\rm GeV}^2\,.
  \label{fb1}
  \end{eqnarray}  
  Note that $s_0^\rho$ coincides with the value quoted in \cite{SVZ},
  while for $b_1$ we get a somewhat larger value than
  Ref.~\cite{GRVW}.  This difference, however, affects the coupling
  only very slightly: with $s^{b_1}_0=2.3$~GeV$^2$ we get
  $f_{b_1}^\perp = 170\,\mbox{\rm MeV}$ with a somewhat worse
  stability.  Note also that it is difficult to specify more precisely
  the value of the continuum threshold $s_0^\rho$: the stability of
  the sum rule does not change much with $s_0^\rho$ in the interval
  (1.3 --1.5) GeV$^2$ (although the value of $f_\rho^\perp$ does),
  which is precisely the disadvantage of having a sum rule of high
  dimension.
  
  Turning to the ``mixed parity'' sum rule (\ref{eq:SRft1}) we first
  note that the contribution of $b_1$ is numerically supressed by the
  exponential factor $\exp[(m_{b_1}^2-m_\rho^2)/M^2]$, so that a
  modest accuracy in $f_{b_1}^\perp$ is sufficient.  Using the value
  in (\ref{fb1}) as input and requiring best stability of the sum rule
  (\ref{eq:SRft1}) by varying $M^2$ and the continuum threshold (see
  Fig.~\ref{fig:mixedparity}a), we get
  \begin{equation}
  f_\rho^\perp = (163\pm 5)\,{\rm MeV},\qquad s_0 = 2.1\, 
{\rm GeV}^2\,.\label{frhot2}
  \end{equation}
  The higher value of $s_0$ (compared to $s_0^\rho$) is expected,
  since the part of the continuum contribution coming from $b_1$ is
  taken into account explicitly on the left-hand side of the sum rule
  (\ref{eq:SRft1}).
  
  On the other hand, since the $b_1$ state is rather wide and its mass
  is very close to the continuum threshold in the pure $1^{--}$
  channel, it is natural to expect that an equally good fit to the sum
  rule can be obtained by ignoring this contribution on the left-hand
  side of (\ref{eq:SRft1}) and fitting the value of the continuum
  threshold to include it effectively.  Remarkably, in this case we
  find a very similar value for the $\rho$ coupling, see
  Fig.~\ref{fig:mixedparity}b:
  \begin{equation}
  f_\rho^\perp = (160\pm 15)\,{\rm MeV},\qquad s_0^\rho = (1.0\pm 0.2)
   \,{\rm GeV}^2\,.
  \label{frhot3}
  \end{equation}
  Note that the sum rule now ``wants'' a much lower value of $s_0$.
  It is instructive to observe that the accuracy is now worse since
  the sum rule remains stable for a rather large interval of $s_0$.
  This is natural, since in this case we do not incorporate an
  additional information about the $b_1$ meson contribution.
  
  To summarize, we find that the positive parity $b_1$ meson
  contributes significantly to the ``mixed parity'' sum rule, but it
  is not possible to separate this contribution from the continuum. In
  effect, the admixture of positive parity states can be described by
  lowering the duality interval for the $\rho$ meson to 1$\,$GeV.  Our
  final result for the $\rho$ meson tensor coupling is
\begin{equation}
f_\rho^\perp = (160\pm 10)\,{\rm MeV}.
\end{equation}
This value is by about 20\% lower than CZ's result \cite{CZreport} and
agrees surprisingly well with an old SU(6) symmetry relation,
$f_\rho^\perp = (f_\pi+f_\rho)/2 \approx 0.17\,$GeV \cite{mallik}.

\begin{figure}
    \centerline{\epsffile{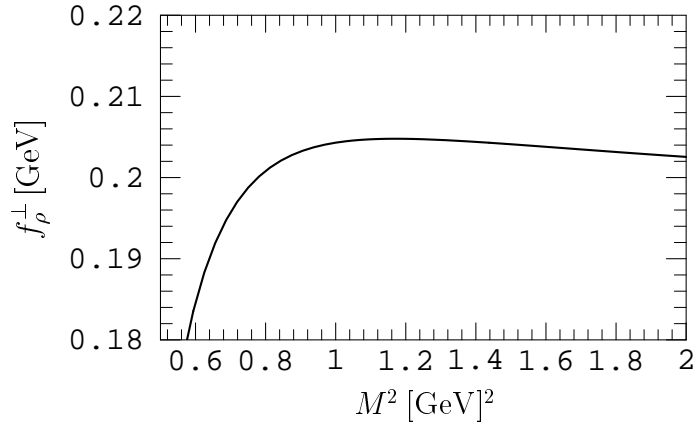}}
  \caption[]{$f_\rho^\perp(1\,{\rm GeV})$ 
    from Eq.~(\protect{\ref{eq:nondiagonal}}) as function of the Borel
    parameter $M^2$ for $s_0=1.5\,{\rm
      GeV}^2$.}\label{fig:nondiagonal}
  \end{figure}

As discussed in \cite{CZZ1,CZreport}, an alternative method to
determine $f_\rho^\perp$ could be to consider the correlation function
of the tensor with the vector current, which is not contaminated by
positive parity states:
  \begin{equation}
   \int\!\! d^4y\, e^{iqy} \,\langle 0|
  T[\bar u(y)\gamma_\mu d(y)\bar d(0)
  \sigma_{\alpha\beta}u(0)]|0\rangle =
  [g_{\alpha\mu}q_\beta-g_{\beta\mu}q_\alpha]\,\chi(q^2)\,.
  \end{equation}
  The correlation function was calculated in \cite{chi,CZZ2} and
  reads:\footnote{The radiative correction to the quark condensate
    contribution is a new result.}
  \begin{equation}
  \chi(q^2) = \frac{2\langle\bar q q\rangle}{q^2}\Bigg\{
  \Big[1-\frac{2\alpha_s}{3\pi}\Big(2\ln\frac{\mu^2}{-q^2}+1\Big)\Big]
  +\frac{m_0^2}{3q^2}+0\cdot \frac{1}{q^4}+\ldots\Bigg\}  
  \end{equation}
  Here $m_0^2\equiv \langle \bar q g\sigma G q\rangle/\langle\bar q
  q\rangle$.  Note that the perturbative contribution vanishes to all
  orders and that the dimension seven operator $\bar q G^2 q$ has zero
  coefficient at tree level \cite{chi}.  The corresponding sum rule
  reads (cf.~\cite{CZZ1,CZreport,chi}):
  \begin{eqnarray}
  e^{-m_\rho^2/M^2} f_\rho^\perp(\mu)\,f_\rho & = &
  -2\langle \bar q q \rangle \left[1+\frac{4}{3}\,\frac{\alpha_s
  }{\pi}\left(\ln\,\frac{M^2}{\mu^2} -\gamma_E -\frac{1}{2} -
  \int\limits_{s_0}^\infty\! \frac{ds}{s}\,e^{-s/M^2}\right)\right.\nonumber\\
  & & \left. {}-\frac{1}{3}\,\frac{m_0^2}{M^2}+
  0\cdot\,\frac{\langle g_s^2G^2\rangle}{M^4}\right],\label{eq:nondiagonal}
  \end{eqnarray}
  and yields $f_\rho^\perp\approx 200\,$MeV as illustrated in
  Fig.~\ref{fig:nondiagonal}. Here we use $f_\rho = 205\,$MeV,
  $\langle \bar q q \rangle(1\,{\rm GeV}) = (-0.25\,{\rm GeV})^3$ and
  $m_0^2=0.65\,{\rm GeV}^2$ at the scale 1~GeV.  The accuracy of this
  sum rule is, however, not competitive to the ones above: the
  uncertainty in the quark condensate alone gives a 10\% error; in
  addition, the study in \cite{chi} indicates possible large
  contributions of excited states to this sum rule, e.g.\ from
  $\rho'(1600)$.  Its significance is, however, that it allows to
  determine the relative {\em sign} of $f_\rho^\perp$ and $f_\rho$,
  which proves to be positive.

  \subsection{Deviations from the Asymptotic Form}
  The deviation of the distribution function from its asymptotic form
  $\phi_\perp(u)\sim u(1-u)$ is quantified by the coefficients $a_n$
  in the expansion (\ref{wf1}). Since the corresponding anomalous
  dimensions are ordered with $n$, one can expect that, at least for
  large scales $\mu$, only a few first terms are important. The QCD
  sum rule approach can be used to estimate $a_2^\perp$.  The
  traditional procedure developed by Chernyak and Zhitnitsky is to
  write down the sum rule for the second moment of the wave function,
  which is related to $a_2^\perp$ by simple algebra:
  \begin{equation}
    \int_0^1 du\,(2u-1)^2 \phi_\perp(u,\mu) = \frac{1}{5} +
  \frac{12}{35}\, a_2^\perp(\mu).
  \label{eq:2}
  \end{equation}
  The corresponding sum rule is obtained from the correlation function
  of the tensor current with the similar operator with two extra
  covariant derivatives $\bar u(y)\sigma_{\mu\xi}x^\xi (i\D\cdot x)^2
  d(y)$.  We find it more appropriate to consider the sum rules
  directly for the coefficients in the expansion in Gegenbauer
  polynomials, which in general correspond to correlation functions of
  the tensor current with the conformal operators
  $$
  \Omega^{T(n)}_\perp(y) = i^n\,(\partial_.)^n \left[\bar
    u(y)\sigma_{\perp .} C^{3/2}_n \left(\frac{\Dr_.-\Dl_.}
      {\Dr_.+\Dl_.}\right) d(y)\right],$$
  where the dots stand for the
  projection on the light-like vector $x_\mu$ and $\partial$ is the
  total derivative.  Note that one of the indices of the $\sigma$
  matrix is projected onto $x_\mu$, while the other one has to be
  taken transverse to the $(x,q)$ plane, where $q$ is the $\rho$ meson
  momentum, see \cite{BF2} for more details.
  
  As a general property of conformal operators \cite{A} the tree-level
  perturbative contribution to the corresponding correlation function
  vanishes (for $n\ne 0$) and the perturbative contribution to the
  corresponding sum rule starts with order $O(\alpha_s)$.  As a
  result, these sum rules are necessarily less stable than the sum
  rules for moments, and their accuracy is seemingly worse. The better
  accuracy of the sum rules for moments is, however, completely
  illusory since in this case the major contribution comes from the
  trivial first term in (\ref{eq:2}), corresponding to the asymptotic
  distribution function, and the contribution of interest is
  numerically suppressed.  Since we should not expect good stability
  for the sum rule for $a_2$, we evaluate this sum rule using
  precisely the same values of the continuum threshold and the same
  ``window'' of the Borel parameter as in the sum rules for
  $f_\rho^\perp$. The instability of the sum rule then gives an
  estimate of the accuracy of the result.\footnote{It has become a
    common practice to choose different values of the continuum
    threshold in the sum rules for different moments. To our point of
    view, the higher fitted values of $s_0$ for higher moments
    $n=2,4,\ldots$ generally reflect the increase of the overall mass
    scale in the correlation function, due to the increasing
    contribution of higher resonances. This rise has nothing to do
    with the change of the interval of duality for the $\rho$ meson
    contribution of interest, which is in fact more likely to {\em
      decrease}.}
  
  It is important to note that the necessity to separate the
  contribution of leading twist does not allow for the separation of
  contributions of opposite parity in the diagonal sum rules. Indeed,
  one may try to start from the correlation function like the one in
  (\ref{eq:decompose}) with two open Lorentz indices (and with the
  substitution of one of the tensor currents by
  $\Omega^{T(n)}_\mu(y)$), and try to isolate the negative parity
  contribution by taking the projection $q^\mu q^\nu
  \Pi^{T(n)}_{\mu\nu} = (qx)^{n+2} \Pi^{-T(n)}(q^2)$.  However, the
  same projection for the defining Eq.~(\ref{def1}) vanishes
  identically since to leading twist accuracy one must put
  contributions of order $q^2=m_\rho^2$ to zero. Thus, this projection
  is in fact saturated by higher twist contributions and is irrelevant
  for our analysis.  Therefore, one cannot get rid of the contribution
  of states with positive parity and a more convenient correlation
  function to consider is: \cite{CZZ2,CZreport}
  \begin{equation}
  i\!\int\!\! d^4y\, e^{iqy} \,\langle 0|
  T[\bar u(y)\sigma_{\mu\xi}x^\xi d(y)\bar d(0)
  \sigma^{\mu\xi}x_\xi (i\D\cdot x)^n u(0)]|0\rangle = 
  -2(qx)^{n+2}\Pi^{T(n)}(q^2).
  \label{eq:CZcorrelator}
  \end{equation}
  It is easy to check that the trace over Lorentz indices picks up the
  required transverse components.
  
  The complete results for the sum rules for the coefficients in the
  Gegenbauer expansion for arbitrary $n$ are given in App.~B.  Note
  that in this case the mass scale in the correlation functions rises
  as $M^2\sim n^2$ for large $n$ as compared to the increase $M^2\sim
  n$ for the moments. This makes the sum rule approach essentially
  useless for the evaluation of $a_n$ with $n>2$.  For the particular
  case $n=2$ we get, using the correlation function
  (\ref{eq:CZcorrelator}):
  \begin{eqnarray}
  \lefteqn{e^{-m_\rho^2/M^2} (f_\rho^\perp)^2(\mu)\, \frac{18}{7}
    \,a_2^\perp(\mu) \ +\mbox{\rm $b_1$ meson}\ =\ }
  \nonumber\\ 
  &=&\frac{1}{2\pi^2}\,\frac{\alpha_s(\mu)}{\pi}\,M^2 
  [1-e^{-s_0/M^2}]\cdot\frac{2}{5}
  +\frac{1}{3 M^2}\,
  \langle\,\frac{\alpha_s}{\pi}\,G^2\,\rangle\, +
  \frac{64\pi}{9M^4}\,\langle\,\sqrt{\alpha_s}\bar q q\,\rangle^2.
  \label{eq:SRa2t}
  \end{eqnarray}
  This sum rule is equivalent to the sum rule for the second moment
  considered in \cite{CZZ1,CZreport}
  \begin{eqnarray}
   \lefteqn{e^{-m_\rho^2/M^2} (f_\rho^\perp)^2(\mu)\,
  \int_0^1\!\! du\,(2u-1)^2 \phi_\perp(u,\mu) 
  +\mbox{\rm $b_1$ meson}\ =}\nonumber\\ 
  &=&\frac{1}{20\pi^2}\int_0^{s_0}\!\! ds\,e^{-s/M^2}
  \left\{ 1 + \frac{\alpha_s}{\pi}\left(\frac{59}{15} + 2 \ln \frac{s}{\mu^2}
  \right)\right\}+\frac{1}{36M^2}\,\langle\frac{\alpha_s}{\pi}\,G^2\rangle + 
  \frac{64\pi}{81M^4}\,\langle\sqrt{\alpha_s}\bar q q\rangle^2.
  \makebox[1.2cm]{\ }
  \label{eq:SRmom2}
  \end{eqnarray}
  provided one takes the same value of the continuum threshold as in
  the sum rule for the tensor coupling (\ref{eq:SRft1}).  Note that
  the sign of the contribution of the four-quark condensate is
  opposite to the result of \cite{CZZ1,CZreport}.\footnote{We have
    recalculated this contribution and get the opposite sign for all
    moments, see App.~B. For the case $n=0$ our result agrees with
    \cite{GRVW}.}

\begin{figure}
  \centerline{\epsffile{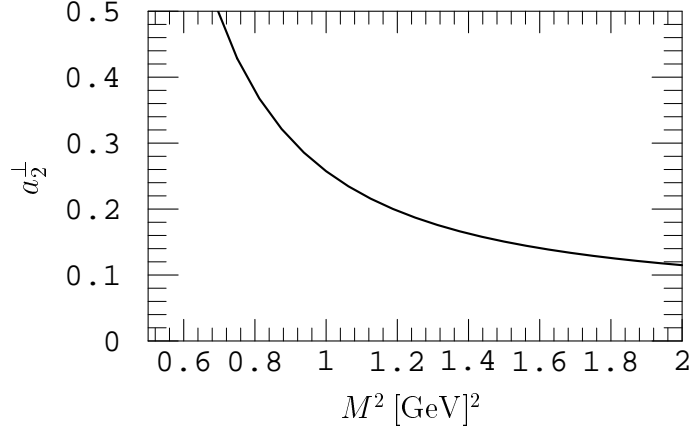}}
\caption[]{$a_2^\perp(1\,{\rm GeV})$ from Eq.~(\protect{\ref{eq:SRa2t}}) as
  function of the Borel parameter $M^2$ for $s_0 =
  1\,$GeV$^2$.}\label{fig:a2t}
\end{figure}
The value of $a_2^\perp$ that follows from the sum rule
(\ref{eq:SRa2t}) is plotted as a function of the Borel parameter in
Fig.~\ref{fig:a2t}.  Note that we do not have an independent estimate
for the contribution of the $b_1$ meson in this case, so we neglect it
and take a low value for the continuum threshold, $s_0=1$ GeV$^2$, on
the right-hand side.  {}From this we get as our final result
  \begin{equation}
    a_2^\perp(\mu=1\,{\rm GeV}) = 0.2\pm 0.1.
  \label{eq:a2t-result}
  \end{equation}
  This has to be compared with $a_2^\perp(\mu=1\,{\rm GeV}) =-0.17$
  from \cite{CZZ1,CZreport}; the difference in sign is mainly due to
  the opposite sign in the contribution of the four-fermion operators
  in \cite{CZZ1,CZreport}.
  
  We have investigated whether adding the $b_1$ contribution as a free
  parameter and requiring best stability in the range $1<M^2<1.5$
  GeV$^2$ could change the result.  We have also tried to follow the
  standard procedure to use the sum rule (\ref{eq:SRmom2}) with $s_0$
  fitted to get best stability.  In both fits the value of $a_2^\perp$
  tends to increase by some (30-50)\%, but we do not find this
  evidence significant enough to influence our estimate.
  
  To avoid an admixture of positive parity states, one can consider,
  instead of (\ref{eq:CZcorrelator}), the correlation function
  \begin{equation}
   \int\!\! d^4y\, e^{iqy} \,\langle 0|
  T[\bar u(y)\gamma_\mu d(y)\bar d(0)
  \sigma_{\alpha\beta}x^\beta (i\D\cdot x)^n)u(0)]|0\rangle =
  [g_{\alpha\mu}(qx)-x_\mu q_\alpha]\,(qx)^n\,\chi^{(n)}(q^2).
  \label{eq:nondiagonal-n}
  \end{equation} 
  The results for $\chi^{(n)}(q^2)$ are available from \cite{BBK}.
  The corresponding sum rule for $a_2^\perp$ reads
  \begin{eqnarray}
  e^{-m_\rho^2/M^2} f_\rho^\perp(\mu)\,f_\rho\,a_2^\perp(\mu) & = &
  -\frac{14}{3}\,\langle \bar q q \rangle \left[1+\frac{29}{18}\,\frac{\alpha_s
  }{\pi}\left(\ln\,\frac{M^2}{\mu^2} -\gamma_E + {\rm const.} -
  \int\limits_{s_0}^\infty\! \frac{ds}{s}\,e^{-s/M^2}\right)\right.\nonumber\\
  & & \left. {}-2\,\frac{m_0^2}{M^2}+
  \frac{85}{216}\,\frac{\langle g_s^2G^2\rangle}{M^4}\right],
  \end{eqnarray}
  where we used vacuum saturation for the contribution of dimension
  seven.  The constant in the radiative correction to the quark
  condensate contribution is not calculated yet.  Unfortunately, due
  to the large coefficient in front of the contribution of the mixed
  condensate, its contribution almost identically cancels the leading
  quark condensate contribution, and the answer depends crucially on
  the contribution of dimension seven, which is poorly known (it is
  suspected that vacuum saturation is strongly violated in this case).
  Thus, from this sum rule one can only get a rough estimate
  $|a_2^\perp|< 0.5$.


\section{Longitudinally Polarized $\rho$ Mesons}
\begin{figure}
  \centerline{\epsffile{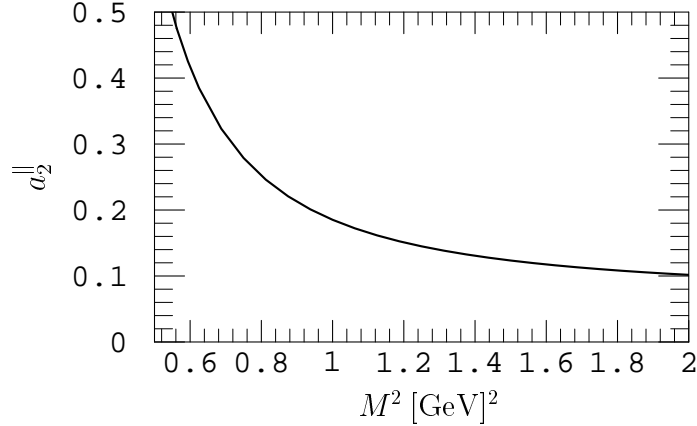}}
\caption[]{$a_2^\parallel(1\,{\rm GeV})$ from 
  Eq.~(\protect{\ref{eq:SRa2l}}) as function of the Borel parameter
  $M^2$ for $s_0=1.5\,{\rm GeV}^2$.}\label{fig:a2l}
\end{figure}

\setcounter{equation}{0} Since the decay constant $f_\rho$ is measured
experimentally (we use the average value $f_\rho = (205\pm 10)\,$MeV
in the numerical analysis), we only need an estimate of the
coefficient $a_2^\parallel$ describing the deviation of the
distribution $\phi_\parallel$ from its asymptotic form.  The
corresponding QCD sum rule calculation has been done by Chernyak and
Zhitnitsky in Ref.~\cite{CZ1}. We update this calculation by taking
into account radiative $O(\alpha_s)$ corrections and using an
up-to-date value of the strong coupling that is slightly larger than
the value used in Ref.~\cite{CZ1}.  The radiative corrections can be
extracted from a paper by Gorskii \cite{gorski}, where he calculated
the correlation function of two axial vector currents (with extra
derivatives), which in perturbation theory and for massless quarks
coincides with the vector correlation function. The complete results
for arbitrary moments are given in App.~B. For $n=2$ we get the sum
rule
\begin{equation}
e^{-m_\rho^2/M^2} f_\rho^2\, \frac{18}{7}
  \,a_2^\parallel(\mu) =\frac{1}{4\pi^2}\,\frac{\alpha_s(\mu)}{\pi}\,M^2 
[1-e^{-s_0/M^2}]
+\frac{1}{2 M^2}\,
\langle\,\frac{\alpha_s}{\pi}\,G^2\,\rangle\, +
\frac{32\pi}{9M^4}\,\langle\,\sqrt{\alpha_s}\bar q q\,\rangle^2\,,
\label{eq:SRa2l}
\end{equation}
which is equivalent to the sum rule for the second moment considered
in \cite{CZ1}:
\begin{eqnarray}
 \lefteqn{e^{-m_\rho^2/M^2} f_\rho^2\,
\int_0^1\!\! du\,(2u-1)^2 \phi_\parallel(u,\mu) 
 =}\nonumber\\ 
&=&\frac{1}{20\pi^2}\,\left(1+\frac{5}{3}\frac{\alpha_s}{\pi}\right)
M^2\left(1-e^{-s_0/M^2}\right)
+\frac{1}{12M^2}\,\langle\frac{\alpha_s}{\pi}\,G^2\rangle + 
\frac{16\pi}{81M^4}\,\langle\sqrt{\alpha_s}\bar q q\rangle^2.
\makebox[0.8cm]{\ }
\label{eq:SRlongmom2}
\end{eqnarray}
With the same input parameters as in Sec.~3, the numerical analysis
yields (see Fig.~\ref{fig:a2l}):
\begin{equation}
    a_2^\parallel(\mu=1\,{\rm GeV}) = 0.18\pm 0.10.
\label{eq:a2l-result} 
\end{equation}
This is in perfect agreement with the original estimate
$a_2^\parallel(\mu =1.1\,{\rm GeV}) \simeq 0.18$ \cite{CZ1}. It also
coincides within the errors with our result for $a^\perp_2$,
Eq.~(\ref{eq:a2t-result}), which means that the distribution
amplitudes $\phi_\parallel$ and $\phi_\perp$ are similar.
 

\section{Summary and Conclusions}
\setcounter{equation}{0} Extending earlier studies
\cite{CZ1,CZZ1,CZreport}, we have performed a re-analysis of $\rho$
meson quark-antiquark light-cone distribution amplitudes of leading
twist. In general, their complete set consists of four independent
functions, but we have shown that to our (twist 2) accuracy the
distributions of transversely polarized quarks in the longitudinally
polarized $\rho$ mesons can be related to the distributions of
longitudinally polarized quarks. The theoretical status of these
relations is identical to the status of the Wandzura-Wilczek relation
\cite{WW} between the polarized structure functions of the nucleon,
$g_1(x,Q^2)$ and $g_2(x,Q^2)$.

\begin{figure}[tb]
  \centerline{\epsffile{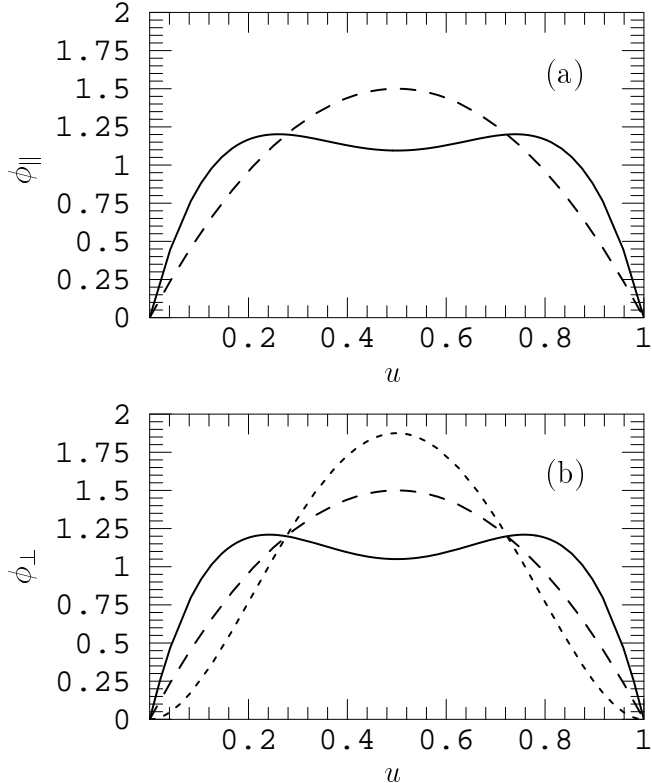}}
\caption[]{Final results for the wave functions $\phi_\parallel$ (a) and 
  $\phi_\perp$ (b) at $\mu = 1\,$GeV (solid lines). Long dashes:
  asymptotic wave functions, short dashes: $\phi_\perp$ according to
  CZ \protect{\cite{CZreport}}.}\label{fig:WF}
\end{figure}

We have given a detailed re-analysis of the QCD sum rules for the
first two moments of the distribution amplitudes, complementing
existing sum rules by the calculation of radiative corrections.  Our
final results for the distribution amplitudes of quarks in
longitudinally polarized and transversely polarized $\rho$ mesons are
shown in Fig.~\ref{fig:WF}a and b, respectively.  The solid curves
correspond to the distributions calculated using the parameters
specified in the text, the dashed curves show the asymptotic
distributions.

We deviate from the results of \cite{CZZ1,CZreport} mainly in the
shape of the distribution amplitude for the transversely polarized
$\rho$ meson, see the short-dashed curve in Fig.~\ref{fig:WF}b.  We
find that the distributions in longitudinally and transversely
polarized $\rho$ mesons coincide to our accuracy, whereas in
\cite{CZZ1,CZreport} a significant difference has been claimed.  This
contradiction is largely due to an opposite sign of the contribution
of four-fermion operators in the corresponding sum rule.  We note that
the sign as given in \cite{CZZ1,CZreport} also contradicts an
independent calculation in Ref.~\cite{GRVW}, which apparently remained
unnoticed. One more consequence of this sign difference is that our
result for the tensor coupling $f_\rho^\perp$ is 20\% lower than in
\cite{CZZ1,CZreport}.

A discussion of phenomenological consequences of our results goes
beyond the tasks of this paper. Since in hard exclusive processes one
typically deals with integrals over quark distributions of type
$$f_\rho \int_0^1 du\,\frac{\phi (u,Q)}{u(1-u)} =
6f_\rho\,[1+a_2(Q)+\ldots]\,,$$
the change in shape of the transverse
$\rho$ distribution suggested by the results of this paper may
increase the rate of the production of transversely polarized $\rho$
mesons by a factor two.  The consequences for exclusive semileptonic
and radiative B decays will be considered in a separate publication
\cite{prep}.

\vspace*{1cm}

\noindent
{\bf Acknowledgements}: We gratefully acknowledge the kind hospitality
of the DESY Theory Group, where this work was finished. We also would
like to thank V.~Chernyak for correspondance.

\vspace*{0.5cm}

\appendix
\setcounter{equation}{0}
\renewcommand{\theequation}{\Alph{section}.\arabic{equation}}


\section*{Appendix A: Transverse Spin Distributions}
\setcounter{equation}{0} \addtocounter{section}{1}

The derivation of relations between longitudinal and transverse quark
spin distributions in the longitudinally polarized $\rho$ meson is in
principle straightforward and can be done similarly to the classical
Wandzura-Wilczek analysis for polarized leptoproduction \cite{WW}.
The major difference is that one must include operators with total
derivatives and that higher twist operators corresponding to total
derivatives of lower twist operators cannot be neglected.

It is convenient to consider the relevant non-local operator at
symmetric quark-antiquark separations:
\begin{equation}
\bar u(-x)\gamma_\mu d(x) =\sum_n x^{\mu_1}\ldots x^{\mu_n}
\frac{1}{n!}\bar u(0)\D_{\mu_1}\ldots \D_{\mu_n}\gamma_\mu d(0)\,.
\end{equation}
Since $x^2=0$ the arising local operators are traceless (contractions
of the type $g_{\mu\mu_k}$ vanish by the equations of motion), but not
fully symmetric in Lorentz indices because of the distinguished index
$\mu$. Therefore, they contain a mixture of contributions of twist~2
and twist~3, which have to be separated:
\begin{equation}
 \bar u(-x)\gamma_\mu d(x) = \Big[\bar u(-x)\gamma_\mu d(x)\Big]_{{\rm
   twist}\, 2}
+ \Big[\bar u(-x)\gamma_\mu d(x)\Big]_{{\rm twist}\, 3}.
\end{equation}
The leading twist 2 contribution by definition contains contributions
of symmetrized operators:
\begin{eqnarray}
\lefteqn{\Big[\bar u(-x)\gamma_\mu d(x)\Big]_{{\rm twist}\, 2} \equiv}
\nonumber\\
&\equiv &
\sum_{n=0}^\infty \frac{x^{\mu_1}\ldots x^{\mu_n}}{n!}
\bar u(0)\Bigg\{\frac{1}{n+1}\D_{\mu_1}\ldots \D_{\mu_n}\gamma_\mu 
+\frac{n}{n+1}\D_{\mu}\D_{\mu_1}\ldots \D_{\mu_{n-1}}\gamma_{\mu_n}
\Bigg\} d(0)\,.\makebox[1cm]{\ }
\label{defsym}
\end{eqnarray} 
Fortunately, the sum can be re-expressed in terms of a non-local
operator \cite{BB88},
\begin{equation}
 \Big[\bar u(-x)\gamma_\mu d(x)\Big]_{{\rm twist}\,2} =
\int_0^1 dv\,\frac{\partial}{\partial x_\mu}\bar u(-vx)\!\not\!x 
d(v x)\,,
\label{BBformula}
\end{equation}
which is easily verified by expanding. An identical expression is
valid for the non-local operator with an additional $\gamma_5$ in
between the quarks.

Using the equations of motion, the difference $\bar u(-x)\gamma_\mu
d(x)- [\bar u(-x)\gamma_\mu d(x)]_{{\rm twist}\, 2}$ can be written in
terms of operators containing total derivatives and
quark-antiquark-gluon operators, see \cite{BB88}. Neglecting quark
masses, one finds:
\begin{eqnarray}
 \Big[\bar u(-x)\gamma_\mu d(x)\Big]_{{\rm twist}\,3} &=&
{}-g_s\int_0^1 \!du\int_{-u}^u \!dv\,\bar u(-ux)
\Big[
u\tilde G_{\mu\nu}(vx)x^\nu\!\not\!x\gamma_5 -ivG_{\mu\nu}(vx)x^\nu\!\not\!x
\Big]d(ux)
\nonumber\\
&&{}+i\epsilon_{\mu}^{\phantom{\mu}\nu\alpha\beta}\int_0^1 udu\, 
x_\nu\partial_\alpha
\Big[\bar u(-ux)\gamma_\beta\gamma_5 d(ux)\Big]\,,
\nonumber\\
\Big[\bar u(-x)\gamma_\mu\gamma_5 d(x)\Big]_{{\rm twist}\,3} &=&
-g_s\int_0^1 \!du\int_{-u}^u \!dv\,\bar u(-ux)
\Big[
u\tilde G_{\mu\nu}(vx)x^\nu\!\not\!x -ivG_{\mu\nu}(vx)x^\nu\!\not\!x\gamma_5
\Big]d(ux)
\nonumber\\
&&{}+i\epsilon_{\mu}^{\phantom{\mu}\nu\alpha\beta}\int_0^1 udu\, 
x_\nu\partial_\alpha
\Big[\bar u(-ux)\gamma_\beta d(ux)\Big]\,,
\label{twist3}
\end{eqnarray}
where $G_{\mu\nu}$ is the gluon field strength, $\tilde G_{\mu\nu}
=(1/2)\epsilon_{\mu\nu\alpha\beta}G^{\alpha\beta}$, and
$\partial_\alpha$ is the derivative over the total translation:
\begin{equation}
\partial_\alpha\Big[\bar u(-ux)\gamma_\beta d(ux)\Big] \equiv 
\left.\frac{\partial}{\partial y_\alpha}
\Big[\bar u(-ux+y)\gamma_\beta d(ux+y)\Big]\right|_{y\to 0}. 
\end{equation}

Note that (\ref{BBformula}) and (\ref{twist3}) are exact operator
relations.  Taking the matrix element between the vacuum and the
$\rho$ meson state, we get
\begin{eqnarray}
\lefteqn{\langle 0 |\Big[\bar u(-x)\gamma_\mu d(x)\Big]_{{\rm twist}\,2}
|\rho^+(p,\lambda)\rangle = }
\nonumber\\ 
&=&\int_0^1 dv\,\frac{\partial}{\partial x_\mu}\langle 0|\bar u(-vx)\!\not\!x 
d(vx)|\rho^+(p,\lambda)\rangle
\ =\ \!\! \int_0^1 dv\,\frac{\partial}{\partial x_\mu}
 (e^{(\lambda)} x)
f_\rho m_\rho\int_0^1 du\, e^{-i\xi vpx} \phi_\parallel(u)
\nonumber\\
&=& e^{(\lambda)}_\mu
f_\rho m_\rho\int_0^1 dv\int_0^1 du\, e^{-i\xi vpx} \phi_\parallel(u)
-ip_\mu
(e^{(\lambda)} x)
f_\rho m_\rho\int_0^1 dv\,v\!\int_0^1 du\, \xi e^{-i\xi vpx} \phi_\parallel(u)
\nonumber\\
&=& p_\mu\, \frac{(e^{(\lambda)} x)}{(px)}\,
f_\rho m_\rho\!\int_0^1 \!\!du\, e^{-i\xi px} \phi_\parallel(u)
+\left( e^{(\lambda)}_\mu -p_\mu \frac{(e^{(\lambda)} x)}{(px)}
 \right) f_\rho m_\rho\!
 \int_0^1 \!\!\!du\!\!\int_0^1 \!\!dv\,e^{-i\xi vpx}\phi_\parallel(u)\,,
\makebox[0.8cm]{\ }
\label{t2}
\end{eqnarray}
where $\xi\equiv 2u-1$ and to arrive at the last line we have used
\begin{eqnarray}
\int_0^1 dv\,v\int_0^1 du\, \xi\, e^{-i\xi vpx} \phi_\parallel(u) &=&
\frac{i}{px}\int_0^1 dv\,v\int_0^1 du\, 
\frac{\partial}{\partial v} e^{-i\xi vpx} \phi_\parallel(u)
\nonumber\\&=&
\frac{i}{px}\int_0^1 du\,\phi_\parallel(u) \Big[e^{-i\xi px}-\int_0^1 dv\,
e^{-i\xi vpx}\Big].
\end{eqnarray}
Note that the matrix element of the twist 2 operator produces both
Lorentz structures, and hence $g_\perp^{(v)}(u,\mu)$ is nonzero to
this accuracy.

Specific for the kinematics in exclusive processes is the generation
of an additional twist 2 contribution by twist 3 operators
proportional to the total derivative $\partial_\alpha\to -ip_\alpha$,
which would vanish in deep inelastic scattering. Taking the matrix
element for the twist~3 operator in the first of Eqs.~(\ref{twist3})
and neglecting three-particle quark-antiquark-gluon distributions of
twist 3 \cite{CZZ2} we get
\begin{eqnarray}
\lefteqn{\langle 0 |\Big[\bar u(-x)\gamma_\mu d(x)\Big]_{{\rm twist}\,3}
|\rho^+(p,\lambda)\rangle = }
\nonumber\\ 
&=&
-\frac{1}{2}(px)^2
\left( e^{(\lambda)}_\mu -p_\mu \frac{(e^{(\lambda)} x)}{(px)}
 \right) \! f_\rho m_\rho
\int_0^1 v^2 dv
\int_0^1 du\, e^{-i\xi vpx} g_\perp^{(a)}(u,\mu)\,.
\label{t3}
\end{eqnarray}
Since, on the other hand,
\begin{eqnarray}
\langle 0 |\bar u(-x)\gamma_\mu d(x) 
|\rho^+(p,\lambda)\rangle &=& p_\mu \frac{(e^{(\lambda)} x)}{(px)}
f_\rho m_\rho\int_0^1 du\, e^{-i\xi px} \phi_\parallel(u,\mu)
\nonumber\\
 &&
\mbox{}+\left( e^{(\lambda)}_\mu -p_\mu \frac{(e^{(\lambda)} x)}{(px)}
 \right) f_\rho m_\rho
 \int_0^1 du\, e^{-i\xi px} g_\perp^{(v)}(u,\mu)\,,\makebox[0.4cm]{\ }
\label{def22}
\end{eqnarray}
we obtain relations between $g_\perp^{(v)}(u,\mu)$,
$g_\perp^{(a)}(u,\mu)$ and $\phi_\parallel(u,\mu)$ by comparing the
Lorentz structures.  At this stage it is convenient to introduce the
moments
\begin{equation}
M^\parallel_n = \int_0^1 du\, \xi^n \phi_\parallel(u,\mu)\,,
\quad
M^v_n = \int_0^1 du\, \xi^n g_\perp^{(v)}(u,\mu)\,,
\quad
M^a_n = \int_0^1 du\, \xi^n g_\perp^{(a)}(u,\mu)\,.
\end{equation}
Expanding (\ref{t2}), (\ref{t3}), (\ref{def22}) in powers of $(px)$,
we get
\begin{equation}
M^v_n = \frac{1}{2}\frac{n(n-1)}{n+1}M_{n-2}^a +\frac{1}{n+1}M_n^\parallel
\label{rec1}
\end{equation}
Similar manipulations with the axial-vector operator (\ref{def3})
produce one more relation
\begin{equation}
\frac{1}{2}M_n^a =\frac{1}{n+2}M_n^v.
\label{rec2}
\end{equation}
Note that the contribution of the leading twist operator $ \langle 0
|[\bar u(-x)\gamma_\mu\gamma_5 d(x)]_{{\rm twist}\, 2}
|\rho^+(p,\lambda)\rangle$ vanishes identically in this case, and the
answer is generated entirely by twist 3 operators, which are reduced
to total derivatives.

Combining (\ref{rec1}) and (\ref{rec2}) we get a simple recurrence
relation,
\begin{equation}
(n+1)M_n^v = (n-1)M_{n-2}^v +M_n^\parallel\,,
\label{rec3}
\end{equation}
the solution of which yields the first relation in (\ref{WW1}).  The
second one then follows from (\ref{rec2}) after some algebra.


\section*{Appendix B: QCD Sum Rules for Arbitrary Moments}
\setcounter{equation}{0} \addtocounter{section}{1} In this appendix we
collect some more definitions and give the sum rules for the
Gegenbauer moments $a_n$ of the longitudinal and transversal $\rho$
meson distribution amplitudes for arbitrary $n$.

We first relate the $a_n$ to hadronic matrix elements of local
operators. To leading logarithmic accuracy, the relevant
multiplicatively renormalizable operators are:
  \begin{eqnarray}
  \Omega^{V(n)}(y) &=& \sum\limits_{j=0}^n c_{n,j} (ix\partial)^{n-j}
  \bar{u}(y)\!\not\!x\,(ix\D)^jd(y)\,,\nonumber\\
  \Omega_\mu^{T(n)}(y) &=& \sum\limits_{j=0}^n c_{n,j} (ix\partial)^{n-j}
  \bar{u}(y)\sigma_{\mu\nu}x^\nu(ix\D)^jd(y)\,,
  \label{eq:defop}
\end{eqnarray}
where $x_\mu$ is a light-like vector, $\sigma_{\mu\nu}
=(i/2)[\gamma_\mu\gamma_\nu-\gamma_\nu\gamma_\mu] $ and
$\D_\mu=\stackrel{\rightarrow}{\partial}_\mu -
\stackrel{\leftarrow}{\partial}_\mu - 2 i g A^a_\mu(y) \lambda^a/2$.
The $c_{n,k}$ are the coefficients of the Gegenbauer polynomials such
that $C_n^{3/2}(x) = \sum c_{n,k}x^k$.  Expanding the defining
relations for $\phi$, Eq.~(\ref{def1})--(\ref{def3}), around the
light-cone, one finds
  \begin{eqnarray}
  \langle 0| \Omega^{V(n)}(0)|\rho\rangle & = &
  (px)^n\,f_\rho m_\rho (ex)\int\limits_0^1\!\! du\, C^{3/2}_n(2u-1)\,
  \phi_\parallel(u,\mu)\nonumber\\
  & = & (px)^n\,f_\rho m_\rho (ex)
  \,\frac{3(n+1)(n+2)}{2(2n+3)}\, a_n^\parallel(\mu)\,,\nonumber\\
  \langle 0| \Omega^{T(n)}_\mu(0)|\rho\rangle & = &
  (px)^n\,i f_\rho^\perp (e_\mu(px)-p_\mu(e x))
  \int\limits_0^1\!\! du\, C^{3/2}_n(2u-1)\,
  \phi_\perp(u,\mu)\nonumber\\
  & = & (px)^n\,i f_\rho^\perp (e_\mu(px)-p_\mu(e x))
  \,\frac{3(n+1)(n+2)}{2(2n+3)}\, a_n^\perp(\mu).
  \end{eqnarray}
  The QCD sum rules \cite{SVZ} are obtained by matching the
  representation in terms of hadronic states to the operator product
  expansion in the Euclidian region for the correlation functions
  \begin{eqnarray}
(qx)^{n+2} \Pi^{V(n)}(q^2) & = & i\!\int\!\! d^4y\, e^{iqy} 
\,\langle 0|T\Omega^{V(n)}(y)\Omega^{\dagger V(0)}(0)\rangle,\nonumber\\
-2 (qx)^{n+2} \Pi^{T(n)}(q^2) & = & i\!\int\!\! d^4y\, e^{iqy} 
\,\langle 0|T\Omega^{T(n)}_\mu(y)\Omega^{\dagger T(0)\mu}(0)\rangle.
  \label{correlator}
  \end{eqnarray}
  Note that the contraction over $\mu$ in the second relation
  automatically projects onto the transverse component
  $\Omega^{T(n)}_\perp$, which is a conformal invariant operator,
  whereas $\Omega^{T(n)}_\mu$ is not.  We find the following sum rules
  for $a_n^\perp$ (for even $n$):
\begin{eqnarray}
\lefteqn{e^{-m_\rho^2/M^2} (f_\rho^\perp)^2(\mu)\,
  \frac{3(n+1)(n+2)}{2(2n+3)}\,a_n^\perp(\mu)\ =}\nonumber\\
&=&\frac{1}{2\pi^2}\,\frac{\alpha_s(\mu)}{\pi}\,M^2 
[1-e^{-s_0/M^2}]
\int\limits_0^1\!\!du\,u\bar u\,C^{3/2}_n(2u-1)\left(\ln u + \ln \bar
  u + \ln^2\,\frac{u}{\bar u}\right) \nonumber\\
&&{}+\frac{1}{24M^2}\,
\langle\,\frac{\alpha_s}{\pi}\,G^2\,\rangle\,(n^2+3n-2) +
\frac{8\pi}{81M^4}\,\langle\,\sqrt{\alpha_s}\bar q q\,\rangle^2
\,(n-1)(n+1)(n+2)(n+4)\,.\makebox[1cm]{\ }\label{eq:B.4}
\end{eqnarray}
The radiative correction in (\ref{eq:B.4}) is a new result.
Similarly, we obtain for $a_n^\parallel$:
\begin{eqnarray}
\lefteqn{e^{-m_\rho^2/M^2} f_\rho^2\,
  \frac{3(n+1)(n+2)}{2(2n+3)}\,a_n^\parallel(\mu)\ =}\nonumber\\
&=&\frac{3}{4\pi^2}\,\frac{\alpha_s(\mu)}{\pi}\,M^2
[1-e^{-s_0/M^2}]\,r_n^\parallel+\frac{1}{24M^2}\,
\langle\,\frac{\alpha_s}{\pi}\,G^2\,\rangle\,(n+1)(n+2)\nonumber\\
&&{} + \frac{8\pi}{81M^4}\,\langle\,\sqrt{\alpha_s}\bar q q\,\rangle^2
\,(n+1)(n+2)(n^2+3n-7)\,.
\end{eqnarray}
In this case a compact answer for the radiative corrections as in
(\ref{eq:B.4}) is not available, but the $r_n^\parallel$ are related
to the radiative corrections to the axial vector correlation function
(with extra derivatives) and for arbitrary $n$ can be expressed in
terms of the coefficients $A_k'$ calculated in \cite{gorski}:
\begin{equation}
r_n^\parallel = \sum\limits_{k=0}^n c_{n,k}\,\frac{A_k'}{(k+1)(k+3)}\,.
\end{equation}
In particular \addtolength{\arraycolsep}{10pt}
$$
\begin{array}{cccc}
\displaystyle A_0' = 1, & \displaystyle A_2' = \frac{5}{3}, &
\displaystyle A_4' = \frac{59}{27}, &\displaystyle A_6' = 
\frac{353}{135},\\[20pt]
\displaystyle r_0^\parallel = \frac{1}{3}, & \displaystyle
r_2^\parallel = \frac{1}{3}, & \displaystyle r_4^\parallel =
\frac{1}{6}, & \displaystyle r_6^\parallel = \frac{83}{810}.
\end{array}
$$
\addtolength{\arraycolsep}{-10pt} For completeness and for
comparison with \cite{CZreport}, we also give the sum rules for the
moments $\langle\xi^n\rangle = \int\!du\,\xi^n\phi(u,\mu)$:
\begin{eqnarray}
\lefteqn{(f_\rho^\perp)^2(\mu) \langle\xi^n\rangle_\perp(\mu)\,
e^{-m_\rho^2/M^2}
= \frac{3}{2\pi^2}\int\limits_0^{s_0}\!\!ds\!\int\limits_0^1\!\! du\,
e^{-s/M^2}\,u\bar u\, (2u-1)^n\left\{1+\frac{\alpha_s}{3\pi}\left(
6-\frac{\pi^2}{3}+2\ln\frac{s}{\mu^2}\right.\right.}\nonumber\\
&&\left.\left.+ \ln u + \ln \bar u + \ln^2\,\frac{
u}{\bar u}\right)\right\} + \frac{n-1}{n+1}\,\frac{1}{12M^2}\,\langle\,
\frac{\alpha_s}{\pi}\,G^2\,\rangle+\frac{64\pi}{81M^4}\,(n-1)\,
\langle\,\sqrt{\alpha_s}\bar q q\,\rangle^2,\makebox[1.4cm]{\ }
\label{eq:moments}\\
\lefteqn{f_\rho^2\langle\xi^n\rangle_\parallel\,e^{-m_\rho^2/M^2} =
\frac{3}{4\pi^2(n+1)(n+3)}\,\left(1+\frac{\alpha_s}{\pi}\,A_n'\right)
M^2\left(1-e^{-s_0/M^2}\right)}\nonumber\\
&&{} + \frac{1}{12M^2}\,\langle\frac{\alpha_s}{\pi}
G^2\rangle + \frac{16\pi}{81M^4}\,(4n-7)\,\langle\,\sqrt{\alpha_s}\bar q q
\,\rangle^2.
\end{eqnarray} 
Note the difference in sign in the last term in Eq.~(\ref{eq:moments})
with respect to (4.25) in Ref.~\cite{CZreport}.
 

\end{document}